\documentclass[english,leqno]{article}
\usepackage[letterpaper]{geometry}
\usepackage{amsmath,amssymb, amsthm}
\usepackage{graphicx}
\usepackage{enumitem}
\usepackage{todonotes}
\usepackage{wrapfig}
\usepackage{mdframed}
\usepackage[bottom]{footmisc} % Footnote at the bottom of page
\usepackage{thm-restate}

\usepackage{comment}
\usepackage{xcolor}
\usepackage{xspace}
\usepackage{xfrac}
\usepackage[normalem]{ulem}

\definecolor{ForestGreen}{rgb}{0.1333,0.5451,0.1333}
\definecolor{DarkRed}{rgb}{0.8,0,0}
\definecolor{Red}{rgb}{1,0,0}

\usepackage{hyperref}

\usepackage[procnumbered,ruled,vlined,linesnumbered, algo2e]{algorithm2e}

\usepackage{cleveref}

\DontPrintSemicolon
\SetKw{KwAnd}{and}
\SetProcNameSty{textsc}
\SetFuncSty{textsc}
\let\oldnl\nl% Store \nl in \oldnl
\newcommand{\nonl}{\renewcommand{\nl}{\let\nl\oldnl}}% Remove line number for one line

\declaretheorem[numberwithin=section]{theorem}
\declaretheorem[numberlike=theorem]{lemma}

\declaretheorem[numberlike=theorem]{corollary}
\declaretheorem[numberlike=theorem]{definition}
\declaretheorem[numberlike=theorem]{claim}

\begin{document}
\title{\Large A Simple Algorithm for Multiple-Source Shortest Paths in Planar Digraphs}

\author{
Debarati Das\thanks{Department of Computer Science, University of Copenhagen, Basic Algorithms Research Copenhagen (BARC). Emails: \texttt{debaratix710@gmail.com, \{kipouridis,koolooz\}@di.ku.dk}.} \thanks{Debarati Das and Evangelos Kipouridis are supported by VILLUM Investigator Grant 16582, Basic Algorithms Research Copenhagen (BARC), Denmark. \includegraphics[height=0.02\textwidth, width=0.035\textwidth]{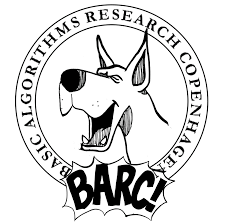}} 
\and
Evangelos Kipouridis\footnotemark[1] \footnotemark[2] \thanks{Evangelos Kipouridis has received funding from the European Union’s Horizon 2020 research and innovation programme under the Marie Skłodowska-Curie grant agreement No 801199. \includegraphics[height=0.02\textwidth, width=0.035\textwidth]{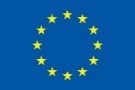}} \and
Maximilian Probst Gutenberg\thanks{ETH Zurich, Email: \texttt{maximilian.probst@inf.ethz.ch}}
\and
Christian Wulff-Nilsen\footnotemark[1] \thanks{Christian Wulff-Nilsen is supported by the Starting Grant 7027-00050B from the Independent Research Fund Denmark under the Sapere Aude research career programme.}
}

\date{}

\maketitle

\begin{abstract} \small\baselineskip=9pt Given an $n$-vertex planar embedded digraph $G$ with non-negative edge weights and a face $f$ of $G$, Klein presented a data structure with $O(n\log n)$ space and preprocessing time which can answer any query $(u,v)$ for the shortest path distance in $G$ from $u$ to $v$ or from $v$ to $u$ in $O(\log n)$ time, provided $u$ is on $f$. This data structure is a key tool in a number of state-of-the-art algorithms and data structures for planar graphs.

Klein's data structure relies on dynamic trees and the persistence technique as well as a highly non-trivial interaction between primal shortest path trees and their duals. The construction of our data structure follows a completely different and in our opinion very simple divide-and-conquer approach that solely relies on Single-Source Shortest Path computations and contractions in the primal graph. Our space and preprocessing time bound is $O(n\log |f|)$ and query time is $O(\log |f|)$ which is an improvement over Klein's data structure when $f$ has small size.
\end{abstract}

\section{Introduction}

In the Planar Multiple-Source Shortest Paths (MSSP) problem, an embedded digraph $G$ is given along with a distinguished face $f$, and the goal is to compute a data structure answering distance queries between any vertex pair $(u,v)$ where either $u$ or $v$ belong to $f$. Data structures for this problem are measured by the \emph{preprocessing time} required to construct the data structure, the \emph{space} required to store the data structure, and the \emph{query time} required to answer a query. 

\paragraph{Applications.} MSSP data structures are a crucial building block for the All-Pairs Shortest Paths (APSP) problem in planar graphs where the data structure needs to be able to return the (approximate or exact) distance between \emph{any} two vertices in the graph. Such data structures are often called \emph{distance oracles} if the query time is subpolynomial in $n$. Broadly, there are three different APSP data structure problems that currently rely on MSSP algorithms:
\begin{itemize}
    \item \uline{Exact Distance Oracles:} In a recent series of breakthroughs, \cite{cohen2017fast,GMWW18,charalampopoulos2019almost,LP20} showed that it is possible to obtain an APSP data structure that requires only space $n^{1+o(1)}$ and query time $n^{o(1)}$ where both~\cite{charalampopoulos2019almost} and the state-of-the-art result in~\cite{LP20} employs MSSP as a building block in their construction.
    \item \uline{Approximate Distance Oracles:} Thorup \cite{thorup2004compact} presented a data structure that returns $(1+\epsilon)$-approximate distance estimates using preprocessing time and space $\tilde{O}(n \epsilon^{-1})$\footnote{We use $\tilde{O}(\cdot)$-notation to suppress logarithmic factors in $n$. To state the bounds in a clean fashion, we assume that the ratio of smallest to largest edge weight in the graph is polynomially bounded in $n$.} and query time $O(\log\log n + \epsilon^{-1})$. Since, his construction time was sped-up by polylogarithmic factors via improvements to the state-of-the-art MSSP data structure \cite{Klein05}.
    \item \uline{Exact (Dynamic) APSP:} A classic algorithm by Fakcharoenphol and Rao \cite{fakcharoenphol2006planar} gives a data structure that uses $\tilde{O}(n)$ space and preprocessing time, and takes query time $\tilde{O}(\sqrt{n})$ to answer queries exactly. A variant of this algorithm further gives a data structure that processes edge weight changes to the graph $G$ in time $\tilde{O}(n^{2/3})$ while still allowing for query time $\tilde{O}(n^{2/3})$. Again, while \cite{fakcharoenphol2006planar} did not directly employ an MSSP data structure, Klein \cite{Klein05} showed that incorporating MSSP leads to a speed-up in the logarithmic factors.
\end{itemize}
We point out that various improvements were made during the last years over these seminal results when considering different trade-offs \cite{fredslund2020truly, long2021planar} or by improving logarithmic or doubly-logarithmic factors \cite{mozes2010shortest, gawrychowski2018improved, mozes2018faster}, however, the fundamental sub-problem of MSSP is present in almost all of these articles. We emphasize that beyond the run-time improvements achieved by MSSP data structure, an additional benefit is a more modular and re-usable design that makes it simpler to understand and implement APSP algorithms.

Another key application of MSSP is the computation of a dense distance graph. An often employed strategy for a planar graph problem is to decompose the embedded planar input graph $G$ into smaller graphs using vertex separators in order to either obtain a recursive decomposition or a flat so called $r$-division of $G$. For each subgraph $R$ obtained, let $\partial R$ denote its set of boundary vertices (vertices incident to edges not in $R$). The dense distance graph of $R$ is the complete graph on $\partial R$ where each edge $(u,v)$ is assigned a weight equal to the shortest path distance from $u$ to $v$ in $R$. Since the recursive or flat decomposition can be done in such a way that $\partial R$ is on a constant number of faces of $R$, MSSP can then be applied to each such face to efficiently obtain the dense distance graph of $R$. There are numerous applications of dense distance graphs, not only for shortest path problems but also for problems related to cuts and flows~\cite{BorradaileKMNW17, Borradaile15GomoryHu, ItalianoFlow11}.

\paragraph{Previous Work.} The MSSP problem was first considered implicitly by Fakcharoenphol and Rao \cite{fakcharoenphol2006planar} who gave a data structure that requires $\tilde{O}(n)$ preprocessing time and space and query time $\tilde{O}(\sqrt{n})$. Since, the problem has been systematically studied by Klein \cite{Klein05} who obtained a data structure with preprocessing time and space $O(n \log n)$ and query time $O(\log n)$. Klein's seminal result was later proven to be tight in $n$ for preprocessing time and space \cite{eisenstat2013linear}. Klein and Eisenstat \cite{eisenstat2013linear} also demonstrated that one can remove all logarithmic factors in the special case of undirected, unit-weighted graphs\footnote{\cite{eisenstat2013linear} also shows how to use this data structure to obtain a linear-time algorithm for the Max Flow problem in planar, unit-weighted digraphs}. Finally, Cabello, Chambers and Erickson \cite{cabello2013multiple} gave an algorithm exploiting the same structural claims as in \cite{Klein05} but give a new perspective by recasting the problem as a parametric shortest paths problem. This allowed them to generalize the result in \cite{Klein05} to surface-embedded graphs of genus $g$, with preprocessing time/space $\tilde{O}(gn \log n)$ and query time $O(\log n)$. 

\paragraph{The Seminal Result by Klein \cite{Klein05}.} On a high-level, the result by Klein \cite{Klein05} is obtained by the observation that moving along a face $f$ with vertices $v_1, v_2, \dots, v_k$, from vertex $v_i$ to $v_{i+1}$, the difference between the shortest path trees $T_{v_i}$ and $T_{v_{i+1}}$ consists on average of $O(n/k)$ edges. \cite{Klein05} therefore suggests to dynamically maintain a tree $T$, initially equal to the shortest path tree $T_{v_1}$ of $v_1$, and then to make the necessary changes to $T$ to obtain the shortest path tree $T_{v_2}$ of $v_2$, and so on for $v_3, \dots, v_k$. Overall, this requires only $O(n)$ changes to the tree $T$ over the entire course of the algorithm while passing through all shortest path trees. 

To implement changes to $T$ efficiently, Klein uses a dynamic tree data structure to represent $T$, and uses duality of planar graphs in the concrete form of an interdigitizing tree/ tree co-tree decomposition, with the dual tree also maintained dynamically as a top tree. Finally, he uses an advanced persistence technique \cite{driscoll1989making} to allow access to the shortest path trees $T_{v_i}$ for any $i$ efficiently. 

Even though formalizing each of these components requires great care, the algorithm by Klein is commonly taught in courses and books on algorithms for planar graphs (see for example \cite{klein2014optimization, demaineCoursePlanarity, topologyErickson}), but with dynamic trees and persistence abstracted to black box components.

\paragraph{Our Contribution.} We give a new approach for the MSSP problem that we believe to be significantly simpler and that matches (and even slightly improves) the time and space bounds of \cite{Klein05}. Our algorithm only uses the primal graph, and consist of an elegant interweaving of Single-Source Shortest Paths (SSSP) computations and contractions in the primal graph. 

Our contribution achieved via two variations of our MSSP algorithm comprises of:
\begin{itemize}
    \item \uline{A Simple, Efficient Data Structure:} We give a MSSP data structure with preprocessing time/space $O(n\log |f|)$ and query time $O(\log |f|)$ which slightly improves the state-of-the-art result by Klein \cite{Klein05} for $|f|$ subpolynomial and otherwise recovers his bounds.
    
    Our result is achieved by implementing SSSP computations via the linear-time algorithm for planar digraphs by Henzinger et al. \cite{HenzingerKRS97}. Abstracting the algorithm \cite{HenzingerKRS97} in black-box fashion, our data structure is significantly simpler than \cite{Klein05} and we believe that it can be taught at an advanced undergraduate level. 
    
    Further, by replacing the black-box from  \cite{HenzingerKRS97} with a standard implementation of Dijkstra's algorithm, our algorithm can easily be taught without any black-box abstractions, at the expense of only an $O(\log n)$-factor to the preprocessing time.
    
    \item \uline{A More Practical Algorithm:} We also believe that our algorithm using Dijkstra's algorithm for SSSP computations is easier to implement and performs significantly better in practice than the algorithm by Klein \cite{Klein05}. We expect this to be the case since dynamic trees and persistence techniques are complicated to implement and incur large constant factors, even in their heavily optimized versions (see \cite{tarjan2010dynamic}). In contrast, it is well-established that Dijkstra's algorithm performs extremely well on real-world graphs, and contractions can be implemented straight-forwardly. 
    
    In fact, one of the currently most successful experimental approaches to compute shortest-paths in road networks is already based on a framework of clever contraction hierarchies and fast SSSP computations (see for example \cite{geisberger2008contraction}), and it is perceivable that our algorithm can be implemented rather easily by adapting the components of this framework.
\end{itemize}

We point out that our result can be shown to be tight in $|f|$ and $n$ by straight-forwardly extending the lower bound in \cite{eisenstat2013linear}. We can report paths in the number of edges plus $O(\log |f|)$ time.

\section{Preliminaries} \label{sec:preliminaries}

Given a graph $H$, we use $V(H)$ to refer to the vertices of $H$, and $E(H)$ to refer to its edges. We denote by $w_H(e)$ or $w_H(u,v)$ the weight of edge $e=(u,v)$ in $H$, by $d_H(u,v)$ the shortest distance from $u$ to $v$ in $H$ and by $P_H[u,v]$ a shortest path from $u$ to $v$. By SSSP tree from a vertex $u\in V(H)$, we refer to the shortest path tree from $u$ in $H$ obtained by taking the union of all shortest paths starting in $u$ (where we assume shortest paths satisfy the subpath property). We use $T(r)$ to denote a tree rooted at a vertex $r$. For a vertex $v\neq r$ of $T$, we let $\pi_T(v)$ denote the parent of $v$ in $T$.

\paragraph{Induced Graph/Contractions.} For a vertex set $X\subseteq V(H)$, we let $H[X]$ denote the subgraph of $H$ induced by $X$. We sometimes abuse notation slightly and identify an edge set $E'$ with the graph having edges $E'$ and the vertex set consisting of endpoints of edges from $E'$. For any edge set $E' \subseteq E(H)$, we let $H / E'$ denote the graph obtained from $H$ by contracting edges in $E'$ where we remove self-loops and retain only the cheapest edge between each vertex pair (breaking ties arbitrarily). If $E$ only contains a single edge $(u,v)$, we slightly abuse notation and write $H/(u,v)$ instead of $H/\{(u,v)\}$. When we contract components of vertices $x_1, x_2, \dots, x_k$ into a super-vertex, we will identify the component with some vertex $x_i$. We use the convention that when we refer to some vertex $x_j$ from the original graph in the context of the contracted graph, then $x_j$ refers to the identified vertex $x_i$.

\paragraph{Simplifying assumptions.} We let $G = (V,E)$ refer to the input graph and assume that $G$ is a planar embedded graph where the embedding is given by a standard rotation system, meaning that neighbors of a vertex are ordered clockwise around it. We assume that $G$ has unique shortest paths, that $f$ is the infinite face $f_{\infty}$, and that this face is a simple cycle with edges of infinite weight. We let $V_{\infty}$ denote the vertex set of $f_{\infty}$ and let $r_0,r_1,\ldots,r_{|V_{\infty}|-1}$ be the vertices of $V_{\infty}$ in clockwise order, starting from some arbitrary vertex $r_0$.

We assume that no shortest path has an ingoing edge to a vertex of $V_{\infty}$; in particular, infinite-weight edges are not allowed to be used in shortest paths. In addition, we assume that every vertex $r_i\in V_{\infty}$ can reach every vertex $u$ of $V\setminus V_{\infty}$ in $G[(V - V_{\infty}) \cup \{r_i\}]$; in words, there is a path from $r_i$ to $u$ that only intersects $V_{\infty}$ in $r_i$.

With simple transformations, it is easy to show that all assumptions can be ensured with only $O(n)$ additional preprocessing time; see Appendix~\ref{sec:input_assumptions} for details.

\section{High-Level Overview}
To make it easier to understand the formal construction and analysis of our new data structure, we start by giving a high-level overview.

\paragraph{Preprocessing.}
We first focus on the preprocessing step which constructs the data structure using a divide-and-conquer algorithm. A formal description can be found as pseudo-code in Algorithm~\ref{alg:MSSP_Preproc} but here we only give a sketch.

The general subproblem is to compute shortest path trees from roots forming an interval $I$ of vertices along $f_{\infty}$ (the initial interval contains every vertex of $f_{\infty}$). Shortest path trees are computed from the two endpoints of $I$ as well as from the middle vertex of $I$. These three roots split $I$ into two equal-sized sub-intervals on which the algorithm recurses. Since a shortest path tree can be computed in linear time in a planar graph, the naive implementation would thus give an $O(n|f_{\infty}|)$ running time. 

%To obtain an efficient construction, we rely on the following well-known result about shortest path trees from the roots on $f_{\infty}$: for any edge $e$ of $G$, the roots of shortest path trees containing $e$ are consecutive on $f_{\infty}$. We will use the following immediate corollary: for any tree $T$ which is a subgraph of $G$, the roots of shortest path trees containing $T$ are consecutive on $f_{\infty}$.
To obtain an efficient construction, we rely on the following well-known result about shortest path trees from the roots on $f_{\infty}$: given two roots and a vertex $u$, the union of the shortest paths from the roots to $u$ split $f_{\infty}$ in two regions. If the shortest path trees from these two roots share an edge $e=(u,v)$ lying in one of the regions, it holds that all roots in the other region contain $e$ in their shortest path trees as well. An immediate corollary is that if the shortest path trees from the two roots share a tree $T$ rooted at $u$ and lying in one of the regions, then all roots in the other region contain $T$ in their shortest path trees.

%To exploit this result in the divide-and-conquer algorithm, let $I'$ be one of the two sub-intervals of $I$ above and let $T_1$ and $T_2$ be the two shortest path trees computed from the endpoints of $I'$. The intersection of edges of $T_1$ and $T_2$ forms a forest. Consider any tree $T$ in this forest and suppose that $I'$ is contained in the interval of roots whose shortest path trees contain $T$. Then for the recursive call to $I'$, $T$ can be contracted since all shortest path trees computed in that recursive call must contain $T$; see Figure~\ref{fig:planarityIdea} with $T(s)$ playing the role of $T$.
To exploit this result in the divide-and-conquer algorithm, let $I'$ be one of the two sub-intervals of $I$ above and let $T_1$ and $T_2$ be the two shortest path trees computed from the endpoints of $I'$. The intersection of edges of $T_1$ and $T_2$ forms a forest. Consider any tree $T$ in this forest and the two regions defined by the endpoints of $I'$ and the root of $T$, as in the previous paragraph. If $T$ lies on the region not containing the roots in $I'$, then for the recursive call to $I'$, $T$ can be contracted since all shortest path trees computed in that recursive call must contain $T$; see Figure~\ref{fig:planarityIdea} with $T(s)$ playing the role of $T$.

Hence, instead of recursing on $I'$ with the entire graph $G$, the algorithm instead recurses on the graph obtained from $G$ by contracting all trees $T$ satisfying the above condition. To ensure that contractions preserve shortest paths, let $T(s)$ be one of the contracted trees. Then edge weights are modified as follows (see Figure~\ref{fig:contract}):
\begin{itemize}
    \item For every edge ingoing to $T(s)$, its weight is increased to $\infty$ unless its endpoint is the root $s$; this ensures that shortest paths can only enter $T(s)$ through $s$.
    \item For every edge $(u,v)$ outgoing from $T(s)$, its weight is increased by the shortest path distance from $s$ to $u$ in $T(s)$; this ensures that the contraction of $T(s)$ does not decrease shortest path distances.
\end{itemize}

It turns out that this simple preprocessing only requires $O(n\log|f_{\infty}|)$ time. To sketch why this is true, consider any edge $e$ of $G$ and let $I_e$ be the interval of roots of $f_{\infty}$ whose shortest path trees contain $e$. Let us refer to the intervals obtained during the recursion as subproblem intervals. We can now make the following observations:
\begin{itemize}
%    \item $I_e$ is the union of $O(\log|f_{\infty}|)$ subproblem intervals.
    \item If a subproblem interval $I$ is disjoint from $I_e$ then $e$ is not part of any shortest path tree in the recursive call to $I$.
    \item If a subproblem interval $I$ is contained in $I_e$ then since $e$ is contracted, $e$ is also not part of any shortest path tree in the recursive call to $I$.
    \item On each recursion level, only $O(1)$ subproblem intervals partially intersect $I_e$.
\end{itemize}
It follows that $e$ is only part of $O(\log|f_{\infty}|)$ shortest path trees in all recursive calls so the total size of all shortest path trees is $O(n\log|f_{\infty}|)$. The time spent on computing a shortest path tree in a graph $H$ is linear in the size of $H$. By sparsity of simple planar graphs, the size of $H$ is proportional to the number of tree edges. It follows that all shortest path trees can be computed in a total of $O(n\log|f_{\infty}|)$ time.

We argue that the additional work spent in the recursive calls can also be executed within this time bound. For instance, determining whether to contract a tree $T$ in the intersection of two shortest path trees $T_1$ and $T_2$ can be done by looking at the cyclic ordering of the two parent edges of $T_1$ resp.~$T_2$ ingoing to the root $r$ of $T$ and the edges of $T$ emanating from $r$; see Figure~\ref{fig:planarityIdea}. This takes time linear in the size of the current graph over all trees $T$.

\paragraph{Handling a query.}
Efficiently answering a query is now simple, given the above preprocessing. Pseudo-code for the query algorithm can be found in Algorithm~\ref{alg:MSSP_Query} but here we only describe it in words and sketch the analysis.

A query for the shortest path distance from a root $r$ of $f_{\infty}$ to a vertex $u$ is done by following the path down the recursion tree for the subproblem intervals containing $r$. For each such subproblem interval $I$, if $r$ is an endpoint of $I$, the shortest path distance is returned since it is stored in the shortest path tree computed from $r$. Otherwise, let $T(s)$ be the tree that contains $u$ and that is contracted for the next recursive step (this includes the special case where $T(s)$ is the trivial tree consisting of the single vertex $u$). The shortest path from $r$ to $u$ passes through the root $s$ of $T(s)$. We compute the $r$-to-$u$ distance by recursively computing the $r$-to-$s$ distance and adding to it the $s$-to-$u$ distance; the latter is precomputed within the time bound for the preprocessing step above. As each recursive step takes $O(1)$ time, we get a total query time of $O(\log|f_{\infty}|)$.

\section{The MSSP Data Structure}\label{sec:ds}
We now give a formal description and analysis of our MSSP data structure.

%\paragraph{Preprocessing of $G$.} To ease the description of our data structure, we preprocess inputs $G$ and $f$ as follows. We let $b_0, b_1, \dots, b_{|f|-1}$ be the vertices on $f$ ordered by their first appearance in the walk along $f$ in clockwise order starting in an arbitrary vertex $b_0$, such that the rest of $G$ is on the right when moving on the walk. For each $0 \leq i < |f|$, we add a vertex $r_i$, and a zero-weight edge $(r_i, b_i')$ embedded in the region enclosed by $f$ that does not contain the rest of $G$. Finally, we add edges $(r_i, r_{i+1 \mod |f|})$ in a simple cycle $C$ again with infinity weights. We let $f_{\infty}$ be the face associated with the region enclosed by the cycle $C$ that does not contain $G$ and denote by $V_{\infty}$ the vertex set of $f_{\infty}$. Note that by construction $f_{\infty}$ is the new infinite face. We denote the new graph by $G_{\infty}$. We now show how to construct our MSSP data structure on $G_{\infty}$ assuming that all queries are on pairs $(u,v) \in V_{\infty} \times V(G)$. It is not hard to see that any query $(b_i, u)$ to $G$ can then be mapped to a query $(r_i, u)$ on $G_{\infty}$.

\begin{algorithm2e}[!ht]
  \SetKwFunction{MSSP}{MSSP}
  \SetKwFunction{Contract}{Contract}

  \setcounter{AlgoLine}{0}
  \SetKwProg{procedure}{Procedure}{}{}
  
  \nonl \procedure{\MSSP{$I = [i_1,i_2], H_I$}}{
  $i\gets \lfloor\frac{i_1+i_2}2\rfloor$\label{lne:start}\;
  \ForEach{$k \in \{i_1,i_2,i\}$}{
    Compute and store SSSP tree $T_{I,k}$ from $r_k$ in $H_I$ \label{MSSP:Trees}
    }
  \lIf{$i_2 - i_1\leq 1$}{ \Return }
  \ForEach(\label{MSSP:J}){$J = [j_1,j_2]\in\{[i_1,i], [i,i_2]\}$}{
    $H_J\gets H_I$\label{lne:copyGraph}\;
    $E_{shared}\gets E(T_{I,j_1})\cap E(T_{I,j_2})$\label{MSSP:Ec}\;
    Let $\mathcal T$ be the collection of maximal vertex-disjoint trees $T(s)$ rooted at $s$ in $E_{shared}$ such that $\pi_{T_{I,j_1}}(s)\neq\pi_{T_{I,j_2}}(s)$, and for each child $v$ of $s$ in $T(s)$ the edges $ (s,v),(s,\pi_{T_{I,j_1}}(s)),(s,\pi_{T_{I,j_2}}(s))$ are clockwise around $s$\label{MSSP:Ts}\;
    \lForEach{$T(s) \in \mathcal{T}$}{
      $H_J \gets \textsc{Contract}(H_J,T(s),J)$
    }
    \MSSP{$J,H_{J}$}\label{lne:recursiveCall}
  }
  }
  
  \nonl \procedure{\Contract{$H_J, T(s), J$}}{
    \lForEach(\label{MSSP:uInTs}){vertex $u\in T(s)$}{
        $(s_J(u),\delta_J(u)) \leftarrow (s,d_{T(s)}(s,u))$ \tcp*[f]{Global variables} \label{MSSP:global}
        } 
      
      \ForEach{$(u,v)\in E(H_{J})$ with exactly one endpoint in $T(s)$}{
        \lIf{$v\notin T(s)$}{
           $w_{H_{J}}(u,v)\gets w_{H_{J}}(u,v) + \delta_J(u)$ \label{MSSP:EdgeInc} \tcp*[f]{$(u,v)$ outgoing from $T(s)$} 
           }
        \lElseIf{$v\neq s$}{
           $w_{H_{J}}(u,v) \gets \infty$ \label{MSSP:EdgeDel}\tcp*[f]{$(u,v)$ ingoing to $T(s) - \{s\}$}
        }
        }
       Contract $T(s)$ to a vertex in $H_{J}$ and identify it with $s$\label{lne:contractTree}\;
       \Return $H_J$\label{MSSP:contraction}
  }

\caption{The procedure $\textsc{MSSP}$ is given an interval $I=[i_1, i_2]$ and a graph $H_I$ obtained by contracting edges in $G$. $H_I$ contains roots $r_j$ for $j\in I$. The initial call is to $([0,|V_{\infty}|-1], G)$.}
\label{alg:MSSP_Preproc}
\end{algorithm2e}

\begin{figure}[ht]
    \centering
    \includegraphics[width=125pt, height=100pt]{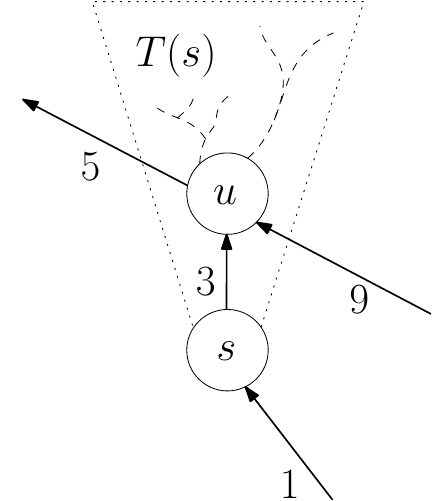}
    \includegraphics[width=125pt, height=100pt]{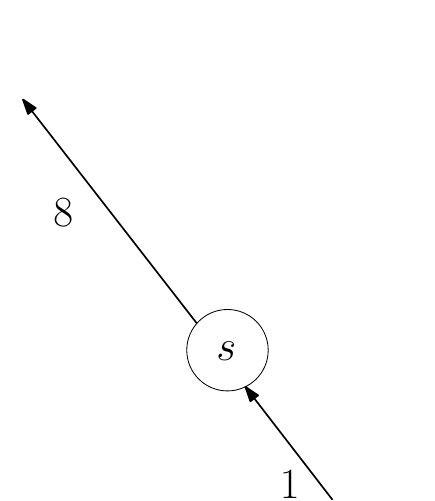}
    \caption{To the left, we have tree $T(s)$. To the right we contract $T(s)$ and identify the supervertex with $s$. We update the weight of outgoing edges and remove the edges ingoing in $T(s)-s$.}
    \label{fig:contract}
\end{figure}

The preprocessing procedure $\textsc{MSSP}(I = [i_1, i_2], H_I)$ in \Cref{alg:MSSP_Preproc} starts by partitioning the interval $I$ into two roughly equally sized subintervals $[i_1, i]$ and $[i, i_2]$, in Lines \ref{lne:start}-\ref{MSSP:Trees}. It then computes the shortest path trees of the boundary vertices $r_{i_1}, r_{i_2}, r_{i}$ in the graph $H_I$ (after removing the other boundary vertices). If $i_2 - i_1 > 1$, the data structure is recursively built for each of the two subintervals $J = [j_1, j_2]$ in the loop starting in \Cref{MSSP:J}. To get the desired preprocessing time, the data structure ensures that the total size of all graphs at a given recursion level is $O(n)$. For each subinterval $J$, this is ensured by letting the graph $H_J$ for the recursive call be a suitable contraction of $H_I$. More precisely, $H_J$ is obtained from $H_I$ by contracting suitable edges $e$ that are guaranteed to be in the SSSP trees of every root $r_j$ with $j\in J$. The construction of $\mathcal{T}$ and the procedure $\textsc{Contract}$ handle the details of these contractions (illustrated in \Cref{fig:contract}). They also store in \Cref{MSSP:global} the information necessary for later queries.

A small but important implementation detail, we point out, is that while the pseudocode specifies that \Cref{alg:MSSP_Preproc} initializes each graph $H_J$ as a copy of $H_I$ in \Cref{lne:copyGraph}, the data structure technically only copies the subset $\{r_{j_1}, r_{j_1+1},\ldots,r_{j_2}\}$ of the vertices on $f_{\infty}$. By our earlier simplifying assumption, the omitted roots are not part of any shortest path tree in any recursive calls involving sub-intervals of $J$ so omitting them will not affect the behaviour of $\textsc{MSSP}$. Including the entire face $f_{\infty}$ in all recursive calls, however, eases the presentation of proofs. 

\paragraph{Query.} The query procedure is straight-forward and given by \Cref{alg:MSSP_Query}. 

\begin{algorithm2e}[ht]
  \SetKwFunction{Query}{Query}

  \setcounter{AlgoLine}{0}
  \SetKwProg{procedure}{Procedure}{}{}
  
  \nonl \procedure{\Query{$u,j,I = [i_1,i_2]$}}{
  \lIf{$j = i_1$ or $j = i_2$}{
    \Return $d_{T_{I,j}}(r_j,u)$} 
   $i\gets \lfloor\frac{i_1+i_2}2\rfloor$\;
  \lIf{$j\leq i$}{
    \Return \Query{${s_{[i_1,i]}(u),j,[i_1,i]}$} $+~ \delta_{[i_1,i]}(u)$}
  \lElse{
    \Return \Query{${s_{[i,i_2]}(u),j,[i,i_2]}$} $+~ \delta_{[i,i_2]}(u)$}
    }
\caption{The procedure to query $d_G(b_j,u)$. Initial call is $\textsc{Query}(u,j,[0,|V_{\infty}|-1])$.}
\label{alg:MSSP_Query}
\end{algorithm2e}

\clearpage\section{Analysis}

We now prove the following theorem which summarizes our main result.

\begin{theorem}\label{Thm:Main}
Let $G$ be an $n$-vertex planar embedded graph and $f_{\infty}$ be the infinite face on $G$. Then we can build a data structure answering the distance $d_G(b_j,u)$ between any vertex $b_j\in f_{\infty}$ and any other vertex $u$ in $O(\log |f_{\infty}|)$ time, using procedure $\textsc{Query}(u,j,[0,|f_{\infty}|-1]))$. Preprocessing requires $O(n \log |f_{\infty}|)$ time and space, using procedure \textsc{MSSP}{$([0,|f_{\infty}|-1],G)$}.
\end{theorem}

\paragraph{Correctness.} Let us first prove correctness of the data structure. We start with the observation that no edge incident to $f_{\infty}$ is ever contracted.

\begin{claim} \label{lemma:infiniteFace}
For any graph $H_I$, no edge incident to $f_{\infty}$ is contracted by \textsc{MSSP}{$(I,H_I)$}.
\end{claim}
\begin{proof}
This is immediate from our assumption that no shortest path has an ingoing edge to a vertex of $f_{\infty}$ since only edges shared by shortest path trees are contracted.
%Observe that any edge that is contracted in some $H_J$ has to be in at least two of the graphs $H_I[(V(H_I)-V_{\infty})\cup \{r_{k}\}]$ for $k\in \{i_1,j,i_2,\}$ and $i_1 < j < i_2$. But since the only edge incident to $V_{\infty}$ in each such graph is the edge $(r_k, b_k)$, no edge on $f_{\infty}$ is contracted.
\end{proof}

\begin{figure}
    \centering
    \includegraphics[width=200pt, height=160pt]{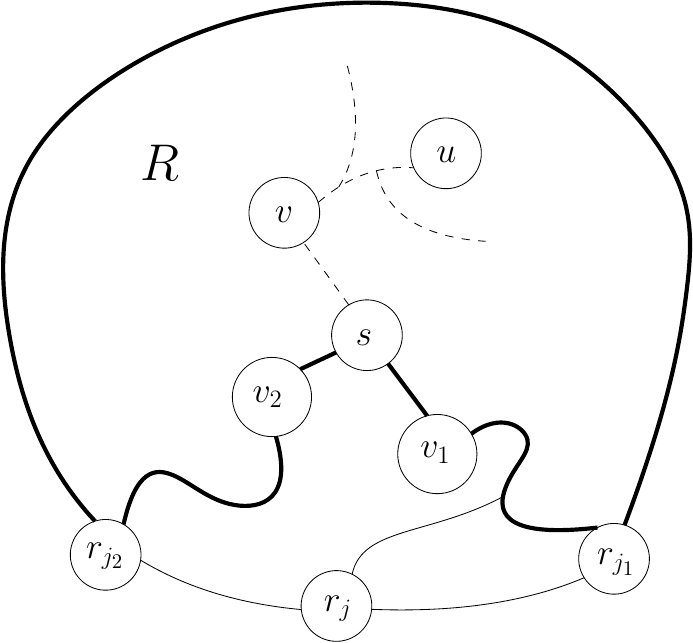}
    \caption{The shortest path trees from $r_{j_1}$ and $r_{j_2}$ share the dashed edges denoting the tree $T(s)$ containing $u$. The shortest $r_{j_1}$-to-$s$, $r_{j_2}$-to-$s$ paths and the subpath from $r_{j_2}$ to $r_{j_1}$ in clockwise order along $f_\infty$ define $C$ (fat line). As $v,v_1,v_2$ are in clockwise order around $s$, $T(s)-s$ is in $R$. For $j\in [j_1,j_2]$, any $r_j$-to-$u$ path intersects either the shortest $r_{j_1}$-to-$s$ or the shortest $r_{j_2}$-to-$s$ path.}
    \label{fig:planarityIdea}
\end{figure}

Next, we prove a lemma and its corollary that roughly show that the contractions made in Procedure~\textsc{MSSP} do not destroy shortest paths from roots on sub-intervals $J$ of $f_{\infty}$. The reader is referred to Figure~\ref{fig:planarityIdea} for intuition. 
We first need the following simple claim.
\begin{claim}\label{clm:thereExistsAPath}
For any invocation of \textsc{MSSP}$(I,H_I)$, any $i \in I$, and any $u \in V(H_I) - V_{\infty}$, some $r_i$-to-$u$ path exists in $H_I[(V(H_I) - V_{\infty}) \cup \{r_i\}]$.
\end{claim}
\begin{proof}
By assumption, an $r_i$-to-$u$ path exists in $G[(V-V_{\infty})\cup\{r_i\}]$. The claim now follows since $H_I$ is obtained from contractions in $G\setminus V_{\infty}$ by \Cref{lemma:infiniteFace}.
%Edge $(r_{i}, b_{i})$ exists in $G$ and is not contractedthe subgraph $G\setminus $ of $G_{\infty}$ is strongly-connected by assumption, and $H_I$ is obtained from contractions in $G_{\infty}$ while preserving $f_{\infty}$ by \Cref{lemma:infiniteFace}.
\end{proof}

\begin{lemma}\label{lemma:PathContain}
Consider any invocation of \textsc{MSSP}{$(I,H_I)$} where $H_I$ has unique shortest paths from $r_i$ for each $i \in I$. Then for each vertex $u \in T(s)\in \mathcal T$ in Line~\ref{MSSP:Ts} and for each $j\in J$, $P_{H_I}[s, u] \subseteq P_{H_I}[r_j, u]$.
\end{lemma}
\begin{proof}
The proof is trivial for $u = s$, thus we assume $u \neq s$. Now consider paths $P_{T_{I,j_1}}[r_{j_1}, s]$ and $P_{T_{I,j_2}}[r_{j_2}, s]$.
Both paths exist by \Cref{clm:thereExistsAPath}. By uniqueness of shortest paths in $H_I$, they share only vertex $s$. 

Next, consider the concatenation $P[j_1, j_2]$ of $P_{T_{I,j_1}}[r_{j_1}, s]$ and the reverse of $P_{T_{I,j_2}}[r_{j_2}, s]$. From the above, $P[j_1,j_2]$ is simple. Further, consider the concatenation $C$ of $P[j_1,j_2]$ and the path segment $F_{\infty}[r_{j_2}, r_{j_1}]$ from $r_{j_2}$ to $r_{j_1}$ in clockwise order around $f_\infty$. We note that $C$ is a cycle. As $f_\infty$ is simple, so is $F_{\infty}[r_{j_2}, r_{j_1}]$, and by our simplifying assumption for shortest paths and \Cref{lemma:infiniteFace}, $P[j_1, j_2]$ only intersects $f_\infty$ in vertices $r_{j_1}$ and $r_{j_2}$. Thus, $C$ is a simple directed cycle.

By the Jordan curve theorem, $C$ partitions the plane into two regions. One region, denoted $R$, is the region to the right when walking along $C$. $T(s)-s$ does not intersect the simple curve $P[j_1, j_2]$ since $H_I$ is a planar embedded graph and since shortest paths are simple. It also does not intersect $F_{\infty}[r_{j_2}, r_{j_1}]$ as there are no ingoing edges to $f_\infty$ in $G_{\infty}$ (and $f_{\infty}$ is preserved in contractions by \Cref{lemma:infiniteFace}). Since for each child $v$ of $s$ in $T(s)$, the edges $(s,v),(s,\pi_{T_{I,j_1}}(s)),(s,\pi_{T_{I,j_2}}(s))$ are clockwise around $s$, children of $s$ are contained in $R$, and hence so is $T(s)-s$.

By the choice of $F_{\infty}[r_{j_2}, r_{j_1}]$, $r_j$ does not belong to $R$ so $P_{H_I}[r_j, u]$ intersects $C$. Since $u\notin F_{\infty}[r_{j_2}, r_{j_1}]$ and since there is no edge $(a,b)$ of $H_I$ with $a\notin V_{\infty}$ and $b\in V_{\infty}$, $P_{H_I}[r_j, u]$ cannot intersect $F_{\infty}[r_{j_2}, r_{j_1}]-\{r_{j_1},r_{j_2}\}$ so it must intersect $C$ in $P[j_1, j_2]$ and hence intersect either $P_{T_{I,j_1}}[r_{j_1}, s]$ or $P_{T_{I,j_2}}[r_{j_2}, s]$; assume the former (the other case is symmetric). Then $P_{H_I}[r_j, u]$ intersects $P_{T_{I,j_1}}[r_{j_1}, s]$ in some vertex $x$. Since shortest paths are unique, the subpath of $P_{H_I}[r_{j}, u]$ from $x$ to $u$ must then equal $P_{T_{I,j_1}}[x, u]$ and since $s$ is on this subpath, the lemma follows.
\end{proof}

\begin{corollary}\label{cor:distancesAreTheSame}
Let $H_J$ be one of the graphs obtained in a call to \textsc{MSSP}{$(I,H_I)$} by contracting edges $E'$ in $H_I$. Then, for each $j \in J$ and each $u \in V(H_J) - V_{\infty}$, $P_{H_J}[r_j, u] = P_{H_I}[r_j, u] / E'$ is unique and $w_{H_J}(P_{H_J}[r_j, u]) = w_{H_I}(P_{H_I}[r_j, u])$.
\end{corollary}
\begin{proof}
By \Cref{lemma:PathContain}, when contracting $P_{H_I}[r_j, u]$ by $E'$, no edge on the path has its weight increased to $\infty$. Further, for any edge $(x,y) \in P_{H_J}[r_j, u]$, either $(x,y)$ existed already in $H_I$ in which case its weight is unchanged, or $(x,y)$ originated from an edge $(w,y)$ for $w \in T(x)$. But in the latter case, $P_{H_I}[r_j,u]$ contains $P_{H_I}[x,w]$ followed by $(w,y)$ (\Cref{lemma:PathContain}), whose weight is $d_{H_I}(x,w)+w_{H_I}(w,y)=d_{H_J}(x,y)$. Thus $w_{H_J}(P_{H_J}[r_j, u]) \leq w_{H_I}(P_{H_I}[r_j, u])$.

It is straight-forward to see that distances in $H_J$ also have not decreased since edges affected by the contractions obtain weights corresponding to paths in $H_I$ between their endpoints or weight $\infty$. Uniqueness follows since two different shortest paths in $H_J$ from $r_j$ to a vertex $u$ would imply two different shortest paths between the same pair in $H_I$ and, by an inductive argument, also in $G$, contradicting our assumption of uniqueness.
\end{proof}

In fact, \Cref{cor:distancesAreTheSame} is all we need to prove correctness of our algorithm.

\begin{lemma}\label{lemma:QueryOutput}
The call \textsc{Query}{$(u,j,I = [i_1,i_2])$} outputs $d_{H_I}(r_j,u)$, for $u \in V(H_I) -V_{\infty}$. In particular, \textsc{Query}{$(u,j,I = [0, |V_{\infty}| - 1])$} outputs $d_{G}(r_j,u)$ for $u \in V$.
\end{lemma}
\begin{proof}
We prove this by induction on $i_2 - i_1$. If $j \in \{i_1,i_2\}$ then \textsc{Query}{$(u,j,[i_1,i_2])$} directly returns $d_{T_{I,j}}(r_j,u)=d_{H_I[(H_I-V_{\infty})\cup \{r_j\}]}(r_j, u)$. This, in turn, is equal to $d_{H_I}(r_j,u)$ as the only $r_j$-to-$u$ paths not considered contain an $\infty$-weight edge with both endpoints in $V_{\infty}$. Therefore the claim holds if $0\leq i_2 - i_1\leq 1$ as $j\in\{i_1,i_2\}$ is implied.

For the inductive step, we thus assume $i_2-i_1>1$ and $i_1<j<i_2$. Let $i=\lfloor \frac{i_1+i_2}2 \rfloor$ and assume $j\le i$ (the case $j > i$ is symmetric). Let $s$ be the vertex in $H_I$ such that $u\in T(s)$, where possibly $u = s$ (Line~\ref{MSSP:uInTs}). By the inductive hypothesis, \textsc{Query}{$(u,j,[i_1,i])$} returns \textsc{Query}{$(s_{[i_1,i]}(u),j,[i_1,i])$}$+\delta_{[i_1,i]}(u)=d_{H_{[i_1,i]}}(r_j,s)+d_{T(s)}(s,u)$. By \Cref{cor:distancesAreTheSame}, $d_{H_{[i_1,i]}}(r_j,s) = d_{H_I}(r_j,s)$. By definition of $T(s)$, $d_{T(s)}(s,u)$ is a suffix of $d_{H_I}(r_{i_1},u)$, meaning that $d_{T(s)}(s,u)=d_{H_I}(s,u)$. Finally, by Lemma~\ref{lemma:PathContain} the shortest path in $H_I$ from $r_j$ to $u$ contains $s$, therefore $d_{H_{[i_1,i],j}}(r_j,s)+d_{T(s)}(s,u) = d_{H_I}(r_j,s) + d_{H_I}(s,u) = d_{H_I}(r_j,u)$.
\end{proof}

\paragraph{Bounding Time and Space.} The following Lemma captures our key insight about \Cref{alg:MSSP_Preproc} that ensures that the algorithm can be implemented efficiently. 

\begin{definition}
Let $\mathcal{I}_h$ be the set of all intervals $I$, such that $\textsc{MSSP}(I, H_I)$ is executed at recursion level $h$ after invoking $\textsc{MSSP}([0, |V_{\infty}|-1], G)$.
\end{definition}
%Note that $\mathcal{I}_h$ is the collection of some intervals $[x_1, x_2], [x_3, x_4], \dots [x_{k-1}, x_{k}] \subseteq [0, |f_{\infty}|-1]$ where $x_j \leq x_{j+1}$ for $j$ even and $x_j < x_{j+1}$ for $j$ odd.

\begin{lemma}\label{lemma:IntervalRecLevel}
For each edge $e = (u,v) \in E(G)$ and recursion level $h$, there are only $O(1)$ intervals $I \in \mathcal{I}_h$ for which there exists an $i \in I$ such that the SSSP tree $T_{I, i}$ from $r_i$ in $H_I$ contains $e$.
\end{lemma}
\begin{proof}
We use induction on $h$. As $\mathcal{I}_0 = \{ [0, |V_{\infty}|-1] \}$ the Lemma is trivial for level $h = 0$. For $h \geq 0$, we show the inductive step $h \mapsto h+1$. Each interval $[i_1, i_2] = I \in \mathcal{I}_h$ satisfies exactly one of the following conditions:
\begin{itemize}
    \item \uline{$\forall i \in I, e \not\in E(T_{I, i})$:} Consider a recursive call $\textsc{MSSP}(J, H_J)$ for $J \subseteq I$ issued in $\textsc{MSSP}(I, H_I)$ where $H_J$ was obtained by contracting some edge set $E'$ in $H_I$. By \Cref{cor:distancesAreTheSame}, for any $j \in J$, $P_{H_J}[r_j, u] = P_{H_I}[r_j, u] / E'$, but due to our condition and $j \in I$, this path cannot contain $e$.
    
    \item \uline{$\forall i\in I, e\in E(T_{I,i})$:} Let $\mathcal I_h'$ be the subset of intervals $I\in\mathcal I_h$ satisfying this condition and let $\mathcal J_{h+1}'$ be the intervals $J\in\mathcal I_{h+1}$ contained in such intervals $I$. We show that for at most one $I\in\mathcal I_h'$ does there exist a sub-interval $J\in\mathcal J_{h+1}'$ where $e$ has not been contracted in $H_J$.
    
    Consider the set $\mathcal P$ of shortest paths in $G$ from each endpoint of an interval in $\mathcal J_{h+1}'$ to vertex $u$. Their union is a tree $T$ with leaves in $V_{\infty}$ and with all edges directed towards the root $u$; see the dashed paths in Figure~\ref{fig:no2CCW}. By definition of $\mathcal I_{h}'$ and $\mathcal J_{h+1}'$ and by repeated applications of Corollary~\ref{cor:distancesAreTheSame} to intervals containing $I$ at recursion levels less than $h$, each shortest path in $\mathcal P$ is a subpath of a shortest path to $v$ in $G$ with $(u,v)$ as the final edge. Since shortest paths are simple, $v$ cannot belong to $T$.
    
    Now, consider an interval $J = [j_1,j_2]\in\mathcal J_{h+1}'$. Let $x$ be the nearest common ancestor of $r_{j_1}$ and $r_{j_2}$ in $T$. Let $C_J$ be the simple directed cycle obtained by the concatenation of the path $T[r_{j_1},x]$, the reverse of the path $T[r_{j_2},x]$, and the path from $r_{j_2}$ to $r_{j_1}$ in counter-clockwise order along $f_{\infty}$. Let $R_J$ be the open region of the plane to the left of $C_J$. Note that shortest path $P_G[x,v]$ is $T[x,u]$ concatenated with $(u,v)$. If $P_G[x,v]$ emanates to the left of $C_J$ at $x$ then since a shortest path cannot cross itself, $v$ must belong to $R_J$. In this case, $v$ cannot also belong to $R_{J'}$ for any other $J'\in\mathcal J_{h+1}'$ since these regions are pairwise disjoint; see Figure~\ref{fig:no2CCW}. Hence, there is at most one choice of $J$ where $P_G[x,v]$ emanates to the left of $C_J$ at $x$.
    
    Using the above notation, it now suffices to show that for an interval $J = [j_1,j_2]\in\mathcal J_{h+1}'$ where $P_G[x,v]$ emanates to the \emph{right} of $C_J$ at $x$, $e$ is contracted in $H_J$.
    
    Let $I\in\mathcal I_h'$ be the parent interval containing $J$. By Corollary~\ref{cor:distancesAreTheSame}, $P_{H_I}[r_{j_1},v]$ is obtained from $P_G[r_{j_1},v]$ by edge contractions and $P_{H_I}[r_{j_2},v]$ is obtained similarly from $P_G[r_{j_2},v]$. Let $(u_1,x)$ resp.~$(u_2,x)$ be the ingoing edge to $x$ in $P_G[r_{j_1},v]$ resp.~$P_G[r_{j_2},v]$. Edge $(u_1,x)$ is not in the shortest path tree from $r_{j_2}$ in $G$ (otherwise, this tree would have both $(u_1,x)$ and $(u_2,x)$ ingoing to $x$) so by Corollary~\ref{cor:distancesAreTheSame}, $(u_1,x)$ is not contracted in $H_I$. Similarly, $(u_2,x)$ is not contracted in $H_I$. It follows that $x$ is a vertex in $H_I$. Let $(x,y)$ be the first edge on $P_{H_I}[x,v]$; this is well-defined since $P_{H_I}[x,v]$ contains at least one edge, namely $e$. By the choice of $J$ and by the embedding-preserving properties of contractions, $(x,y)$, $(x,\pi_{T_{I,j_1}}(x))$, and $(x,\pi_{T_{I,j_2}}(x))$ are clockwise around $x$. Inspecting the description of \textsc{MSSP}{$(I,H_I)$}, we see that $P_{H_I}[x,v]$ is contained in a tree $T(s)\in\mathcal T$ with $s = x$, so $e$ is contracted in $H_J$.
    
    %Consider each iteration of the foreach-loop in \Cref{MSSP:J} in $\textsc{MSSP}(I, H_I)$. Let $J \subseteq I$ be one of the at most two intervals considered during one such iteration.
    
    %Note that if $e \in T(s)$ for some $T(s) \in \mathcal{T}$, then to construct $H_J$ from $H_I$ the edges in $T(s)$ are contracted in  \cref{lne:contractTree}, and therefore $e$ cannot be contained in $H_J$.
    
    %We therefore assume that $e \not\in T(s)$ for any $T(s)\in\mathcal T$. We claim that there is at most one interval $[j_1, j_2] \in \mathcal{I}_{h+1}$ satisfying the condition of the lemma. The proof of this claim is illustrated in \Cref{fig:no2CCW}.
    
    \begin{figure}[ht]
        \centering
        \includegraphics[width=210pt, height=160pt]{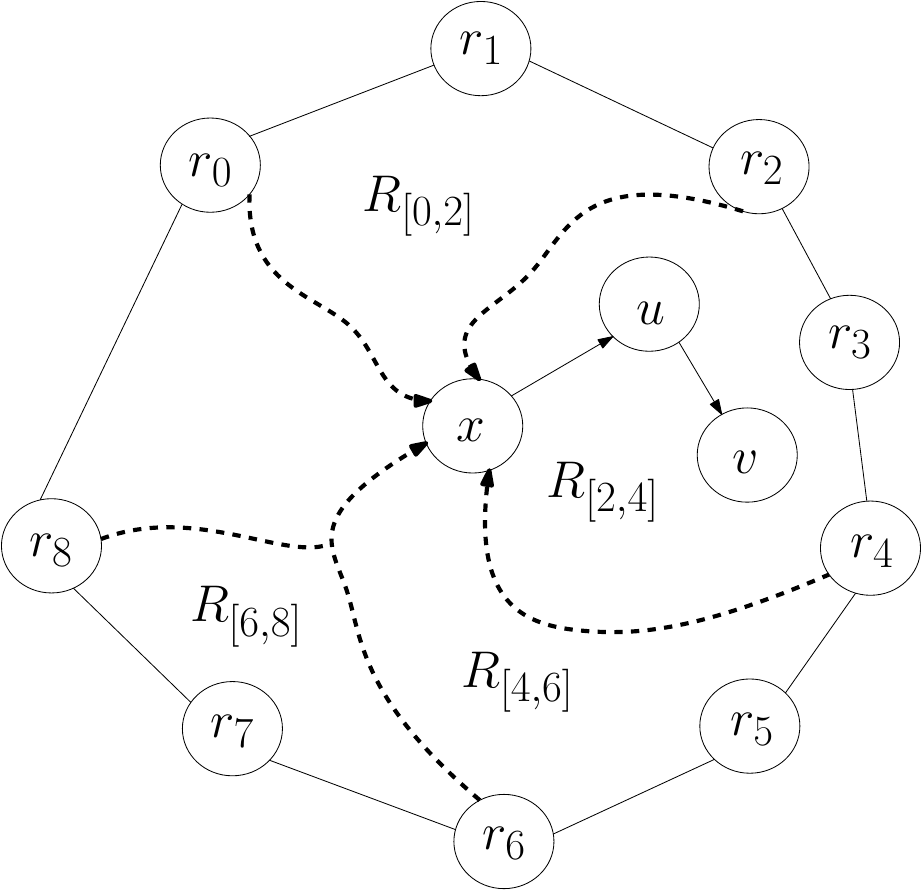}
        \caption{Dashed lines are parts of shortest paths $P_{G}[r_i,v]$ from vertices $r_i$ in $f_{\infty}$ to $v$ and form a tree.
        In the second recursion level we have intervals $I_0=[0,2], I_1=[2,4], I_2=[4,6], I_3=[6,8]$. Vertex $x$ is the nearest common ancestor of $r_2$ and $r_4$ and $R_{[2,4]}$ is the open region to the left of the simple cycle defined by $P_{G}[r_2,x]$, the reverse of $P_{G}[r_4,x]$ and the part of $f_\infty$ from $r_4$ to $r_2$ in counter-clockwise order. Regions $R_{[0,2]},R_
        {[4,6]},R_{[6,8]}$ are defined similarly for the other intervals.
        Focusing on interval $I_1$, let $y$ be the vertex after $x$ towards $u$ in the tree (in this case $y=u$). Edges $(x,y), (x,\pi_{P_{G}[r_2,u]}(x)), (x,\pi_{P_{G}[r_4,u]}(x))$ are in counter-clockwise order, implying that $v$ belongs to $R_{[2,4]}$. Since regions are pairwise disjoint, $v$ belongs to no other region.}
        \label{fig:no2CCW}
    \end{figure}

    \item \uline{$\exists i,i'\in I$, $e\in E(T_{I,i})$ and $e\notin E(T_{I,i'})$:} At most two intervals $I\in\mathcal I_h$ satisfy this condition which can then spawn at most four intervals in $\mathcal I_{h+1}$; this follows from repeated applications of Corollary~\ref{cor:distancesAreTheSame} combined with the easy observation that roots of $f_{\infty}$ whose shortest path trees in $G$ contain $e$ are consecutive in the cyclic ordering along $f_{\infty}$~\cite{Klein05}.
\end{itemize}
\end{proof}

\begin{corollary}
The procedure $\textsc{MSSP}([0, |V_{\infty}|-1], G)$ uses $O(n\log |f|)$ time and space. Procedure \textsc{Query}$(u,j, [0, |V_{\infty}|-1])$ has $O(\log |f|)$ running time.
\end{corollary}
\begin{proof}
For each recursion level $h$, \Cref{lemma:IntervalRecLevel} implies that each edge $e \in E(G)$ appears in at most $O(1)$ computed SSSP trees. Further, $\sum_{I\in\mathcal I_h}|I| = O(n)$. Thus, the total number of vertices in $H_I$ summed over all $I\in\mathcal I_h$ is $O(n)$. By sparsity of planar graphs, the total number of edges in these graphs is $O(n)$ as well.%it follows that $O(\sum_{I \in \mathcal{I}_h} (|H_I| - |V_{\infty}| + |I|)) = O(n)$.

For each such graph $H_I$, each SSSP computation in $H_I$ can be implemented in time linear in the size of $H_I$ by \cite{HenzingerKRS97}. We also spend this amount time on constructing the graphs $H_J$ from $H_I$ since the set $E_{shared} \subseteq E(H_I)$ and associated trees $T(s)$ can be found in $O(1)$ time per edge. Each edge is outgoing from at most one and ingoing to at most one contracted tree and thus possibly changing its weight requires only $O(1)$ time.

It follows that the entire time spent on recursion level $h$ is $O(n)$. Since there are at most $O(\log |f_{\infty}|)$ levels, the first part of the corollary follows. The second part follows since each recursive step in $\textsc{Query}$ takes constant time.
\end{proof}

\bibliographystyle{siam}
\bibliography{apsp}
%\printbibliography%[heading=bibintoc] % Make bibliography show up in table of contents

\appendix

\section{Satisfying the input assumptions}\label{sec:input_assumptions}
In this section, we show how to ensure the input assumptions from the preliminaries.

If $f$ is not the external face $f_{\infty}$, a linear time algorithm can reembed $G$ to ensure this.

Next, we need $f_{\infty}$ to be a simple cycle. We let $b_0, b_1, \dots, b_{|f_{\infty}|-1}$ be the vertices on $f_{\infty}$ ordered by their first appearance in the walk along $f_{\infty}$ in clockwise order starting in an arbitrary vertex $b_0$, such that the rest of $G$ is on the right when moving on the walk. For each $0 \leq i < |f_{\infty}|$, we add a vertex $r_i$, and a zero-weight edge $(r_i, b_i')$ embedded in the region enclosed by $f_{\infty}$ that does not contain the rest of $G$. Finally, we add edges $(r_i, r_{i+1 \mod |f_{\infty}|})$ in a simple cycle, with infinite weights, and redefine $f_{\infty}$ to be this cycle. It is not hard to see that this transformation can be done in $O(n)$ time and that any query $(b_i, u)$ in the original graph can be answered by querying $(r_i, u)$ in the transformed graph.

Let $V_{\infty} = \{r_0,r_1,\ldots,r_{|f_{\infty}|-1}$. To ensure the requirement that every $r_i\in V_{\infty}$ can reach every vertex of $V\setminus V_{\infty}$ in $G[(V - V_{\infty}) \cup \{r_i\}]$, we add suitable edges of large finite weight to $G\setminus V_{\infty}$ to make this subgraph strongly connected without violating the embedding of $G$.

Disallowing shortest paths from having edges ingoing to $V_{\infty}$ will not disallow any shortest path from $b_i$ in the original graph: each such path can be mapped to a path of the same weight in the transformed graph that only intersects $f_{\infty}$ in $r_i$.

Finally, unique shortest paths can be ensured using the deterministic perturbation technique in~\cite{UniqueShortestPaths}).
\end{document}